# Mass asymmetry effects on multifragmentation.

Supriya Goyal and Rajeev K. Puri*
*Department of Physics, Panjab University, Chandigarh-160014, INDIA*
* email: rkpuri@pu.ac.in

## Introduction

Several experimental and theoretical work involving both symmetric as well as asymmetric reactions have been performed during last years [1,2]. The mass asymmetry of the reaction is defined as $\eta = (A_T - A_P)/(A_T + A_P)$; where $A_T$ and $A_P$ are the masses of target and projectile. The $\eta = 0$ corresponds to the symmetric reactions whereas non zero values of $\eta$ corresponds to mass asymmetric reactions. The reaction dynamics in a symmetric reaction is quite different compared to mass asymmetric one. This is due to the fact that in symmetric reactions, the excitation energy leads to larger compression while mass asymmetric reactions lack the compression because a large part of excitation energy is in the form of thermal energy [2]. The asymmetry of a reaction studied experimentally can be as large as 0.89 (in the case of C+Au [3]). Recently, FOPI collaborations have also done study on the mass asymmetric reactions of Ca+Au/Au+Ca [4]. In the literature, a lot of isolated studies are available on the mass asymmetric reactions [2,3,4,5]. The effect on mass asymmetry on the collective flow and its disappearance has also seen in recent work of Puri and collaborations [6]. But the effect of mass asymmetry on the multifragmentation using quantum molecular dynamics (QMD) model is not available anywhere. In the present study, we aim to address this problem using QMD model [7].

## Model

The QMD model simulates the heavy-ion reactions on event by event basis. This is based on a molecular dynamic picture where nucleons interact via two and three-body interactions. The nucleons propagate according to the classical equations of motion:

$$\frac{d\mathbf{r}_i}{dt} = \frac{dH}{d\mathbf{p}_i} \text{ and } \frac{d\mathbf{p}_i}{dt} = -\frac{dH}{d\mathbf{r}_i}, \quad (1)$$

where H stands for the Hamiltonian which is given by

$$H = \sum_i \frac{\mathbf{p}_i^2}{2m_i} + V^{tot}. \quad (2)$$

Our total interaction potential $V^{tot}$ reads as

$$V^{tot} = V^{Loc} + V^{Yuk} + V^{Coul}, \quad (3)$$

where $V^{Loc}$, $V^{Yuk}$, and $V^{Coul}$, respectively, the local (two and three-body) Skyrme, Yukawa, and Coulomb potentials. For details, the reader is referred to Ref. [6].

## Results and discussion

For the present study, we performed simulations for the central reactions of $^{197}Au + ^{197}Au$ ($\eta=0$), $^{129}Xe + ^{197}Au$ ($\eta=0.21$), $^{40}Ar + ^{108}Ag$ ($\eta=0.46$), $^{16}O + ^{80}Br$ ($\eta=0.67$), and $^{16}O + ^{108}Ag$ ($\eta=0.74$) at an incident energy of 200 MeV/nucleon, using soft equation of state along with free energy dependent nucleon-nucleon cross-section. We have employed MSTB(2.1) [8] and SACA(2.1) [9] algorithms for clusterization. The freeze out time is taken to 250 fm/c in case of MSTB(2.1) and $T_{min}$ (at which heaviest fragment ($<A_{max}>$) attains first minima) in case of SACA(2.1). The value of $T_{min}$ varies with change in mass asymmetry as well as energy. In Fig. 1, we display the reduced multiplicities of mass of heaviest fragment ($<A^{max}>$), free nucleons, light charged particles (LCP's) ($2\leq A \leq 4$), and fragments with masses $5\leq A \leq 9$, $5\leq A \leq 48$ as well as $5\leq A \leq 65$. One can clearly see from the figure that the mass asymmetry of a reaction has a significant effect on the dynamics of the reaction. Due to the decrease in number of nucleon-nucleon collisions with increase in mass asymmetry, the size of $<A_{max}>$ increases with increase in η, whereas the multiplicities of free nucleons and various fragments decreases. This is because of decrease in participant region. The decrease in the multiplicity of various fragments with MSTB(2.1) is more pronounced as compared to SACA(2.1). One should also note that for nearly symmetric reactions, both clusterization algorithms gives nearly same results while as the mass asymmetry increases, the results starts differing.

## Acknowledgments

This work is supported by a research grant from the Council of Scientific and Industrial Research (CSIR), Govt. of India, vide grant No. 09/135(0563)/2009-EMR-1.

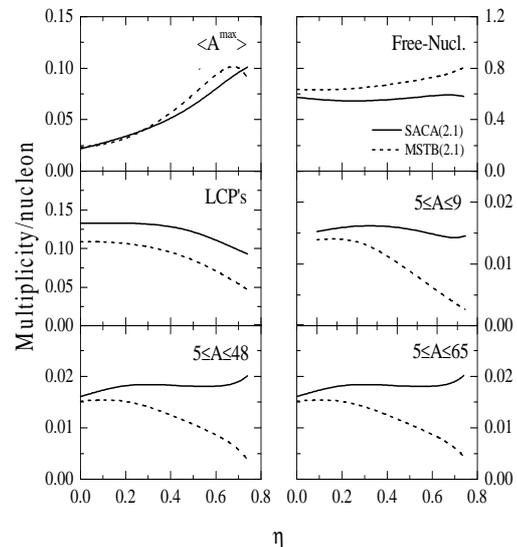

**Fig. 1** Different reduced fragment multiplicities as a function of mass asymmetry of the reaction (η). The results for SACA(2.1) and MSTB(2.1) are represented, respectively, by solid and dashed lines.